\documentstyle[epsfig,twocolumn,prl,aps]{revtex}
\begin{document}

\title{Creating massive entanglement of Bose condensed atoms}
\author{Kristian Helmerson$^{1}$ and Li You$^{2}$}
\address{$^{1}$Atomic Physics Division, National Institute of
Standards and Technology, Gaithersburg, MD 20899-8424\\
$^{2}$School of Physics, Georgia Institute of Technology,
Atlanta, GA 30332-0430}
\date{\today}
\maketitle

\begin{abstract}
We propose a direct, coherent coupling scheme that can
create massively entangled states of Bose-Einstein condensed atoms.
Our idea is based on an effective interaction between two atoms from
coherent Raman processes through a (two atom) molecular
intermediate state. We compare our scheme with other recent proposals for
generation of massive entanglement of Bose condensed atoms.
\end{abstract}


\narrowtext

Entanglement lies at the heart of the difference between
the quantum and classical multi-particle world. It is
a phenomena which facilitates quantum
information and quantum computing with many qubits.
Recently, several interesting developments have occurred
in studies of massively entangled atomic states.
Based on the proposals of M\o lmer and S\o rensen~\cite{mol},
a controlled, entangled state of 4-ions was successfully created
by the NIST ion trap group~\cite{4-ions}.
Zeilinger and coworkers prepared three entangled photon
or Greenberger-Horne-Zeilinger (GHZ) states by selecting from two beams of
entangled photon pairs~\cite{anton}. Entanglement
between two atoms and a microwave photon was also
detected in a ``step-by-step" process~\cite{paris}.

Of these and other related developments, the idea of 
M\o lmer and S\o rensen~\cite{mol} is especially interesting. They
proposed a direct coupling to the multi-particle, entangled
final state through a virtual, intermediate
state which was a common (quantum) mode of the motion of all the ions.
Similar type interactions were also proposed by
Milburn~\cite{mil}. Both proposals allow
for creation of massive entangled states by
unitary evolution, starting from certain pure initial states.

A Bose-Einstein condensate of a dilute atomic vapor 
is a convenient source of atoms, well approximated as
initially being in pure and separable states.  
S\o rensen {\it et al.}~\cite{zoller}
suggested creating massively entangled,
spin squeezed states from a two component condensate
using the inherent atom-atom interactions.
Spin-exchange collisional interactions in a
spinor condensate was also proposed as a candidate for
creating entangled pairs of atoms~\cite{meystre,zoln}.
Most of these proposals work in the two mode approximation
where one motional state is assumed for each spinor
component of condensed atoms. Similar studies were
previously performed for condensate atoms
in a double well potential~\cite{tony,sipe,you,milburn}.

Raman transitions mediated by long-range dipole-dipole interactions 
have been proposed in entanglement schemes for quantum 
computing~\cite{ivan,lukin}.
In this letter, we propose a new type of coupling also based on the two atom, 
effective interaction from a Raman process
through intermediate, molecular states. 
We show that our coupling applied to condensate atoms 
can achieve both massively entanglement, similar to 
that of M\o lmer and S\o rensen~\cite{mol} and improved 
spin squeezing~\cite{zoller,meystre,zoln,ueda}. 

\begin{figure}
\centerline{\psfig{figure=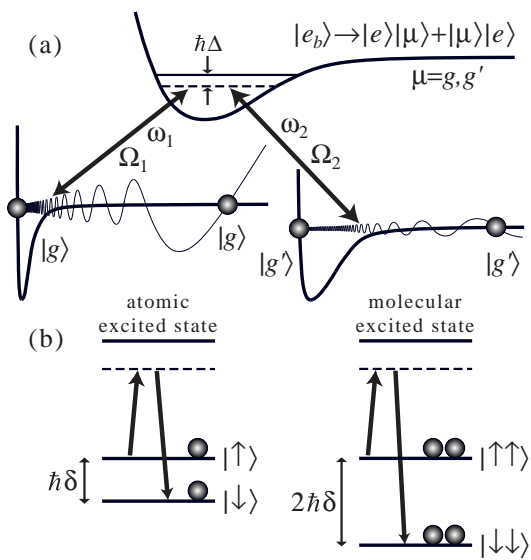,width=2.75in}\\[12pt]}
 \vskip -0.125in
\caption{(a) The general two-photon Raman scheme via an intermediate, 
excited, molecular transition. (b) Raman coupling between two spin 
states illustrated for the atomic and molecular case.  In the atomic 
case, single atoms can be prepared in a coherent superposition of spin 
up and spin down.  In the molecular case, pairs of atoms are prepared 
in a coherent superposition of spin up and spin down.}
\label{fig1}
\end{figure}

Consider a system (see Fig.\ref{fig1}) involving two
$\Lambda$-type atoms whose initial and final states
are described by the same non-interacting atomic states 
$|g\rangle$ and $|g'\rangle$.
Transitions between $|g\rangle$ and $|g'\rangle$ for the pair of 
atoms are coupled by two laser fields (frequency $\omega_{i}$, wavevector $\vec k_{i}, 
k_{i}=\omega_{i}/{\rm c}$, and Rabi frequency $\Omega_{i}$) 
through an intermediate excited state, which is chosen to be
a bound, molecular excited state.

We consider an excited molecular
state $|e_b\rangle$ asymptotically connected, for large internuclear separation, 
to one ground $|\mu \rangle, \mu=g,g'$ and one excited atom $|e\rangle$.
An example is the $0^-_g$ state, which has been 
extensively discussed in the context of 
photo-association of ultracold atoms~\cite{pas,verhaar}. 
The recent experiment by
Heinzen and coworkers on the production of ground state molecules
from an atomic condensate by a two-photon Raman process~\cite{heinzen} 
provides additional motivation to explore our ideas
experimentally. The photo-association process~\cite{pas,verhaar,julienne,matt} 
used in Heinzen's experiment relies
on the transition strength of going from a ground `free' (two atom) to a
ground `bound' (molecular) state via an intermediate, excited `bound' (molecular)
state. On the other hand, what we desire is a transition
from a ground `free' to another ground `free' state via an
intermediate, excited `bound' state. 

The question of whether this is possible or not 
for trapped ultracold cloud of atoms
does not seem to depend on the sample density
(in the weakly interacting limit), but will depend, 
to a large degree, on the trap strength and
the excited, bound state structure.
That is, we can determine the strength of the transition by 
considering the collision of a pair of atoms in the 
combined trap and molecular potentials and then summing 
over the number of pairs of atoms available.

Assuming all trapping potentials for the ground and 
excited states to be harmonic,
there is a separation of the trapping potentials in terms of the 
center-of-mass $\vec R$ and relative
coordinates $\vec r$ of the two atoms.
When short range interactions are
approximated by their optical contact forms,
Wilkens and coworkers provide analytic solutions for 
the center-of-mass and relative motion of two
interacting atoms inside a harmonic trap~\cite{martin}.
In principle we can find all bound states in the $\vec r$ potential 
due to the external trapping potential, even with a realistic molecular
interaction potential~\cite{lowenergy}.
In general one can include all relative motional states
and still be able to find the effective coupling between
the selected electronic states~\cite{youM}. 
We consider, however, only the lowest unbound state of the relative 
motion $|0\rangle$ in the electronic ground state,
a situation well approximated by Bose condensed atoms.

If we choose a state fairly deep in the excited state 
molecular potential, then adjacent molecular states will be well 
separated in energy and we can consider coupling to only one 
intermediate bound state with vibrational quantum number $m_b$.  
Furthermore, for a sufficiently 
large detuning of the coupling lasers from the excited state 
$e_b,|m_b\rangle$, 
we can perform an adiabatic elimination of the excited state.
Neglecting configurations not directly involved in the two photon 
process~\cite{secondorder},
the Hamiltonian for the two atoms initially in the $g$ state can be written as

\begin{eqnarray}
{\cal H}&&=\!\sum_{\mu=g,g'}
\left[{{\vec P}^2\over 2(2M)}+V_{tR}(\vec R)
+2\hbar\omega_{\mu g}+E_{0}\right]\nonumber\\
&&\times
|\mu,\mu;0\rangle\!\langle \mu,\mu;0|+V_R(\vec R,t),
\end{eqnarray}
where
\begin{eqnarray}
V_R={\hbar\Omega_R\over 2}|\eta_{0 m_b}|^{2}
e^{i\Delta\omega t}e^{i\vec K\cdot \vec R}
|\mu ,\mu ;0\rangle\!\langle g,g;0| + h.c.
\end{eqnarray}
$\Delta\omega=\omega_{2}-\omega_{1}$, $\vec K=\vec k_{1}-\vec 
k_{2}$, $V_{tR}$ is the trapping potential for the center-of-mass 
motion, $\hbar \omega_{\mu g}$ is the energy difference between the 
atomic states $g$ and $g$ or $g'$ and, for simplicity, we have assumed 
the same relative motional state $|0\rangle$, with energy $E_{0}$, 
for the pair of atoms in $g$ or $g'$ (i.e. atoms in 
state $g$ or $g'$ see the same trapping potential). 
Since $|e_b\rangle$ is asymptotically connected with
$|e\rangle|\mu\rangle+|\mu\rangle|e\rangle$
(for $\mu=g,g'$) we can set the dipole matrix element 
$\vec d_i =\vec d_i' =\vec d$, for the
$|\mu\rangle$ to $|e\rangle$ transition of atom $i$. Then 
$\Omega_{R}=\Omega_{1}\Omega^{*}_{2}/2\Delta$ is simply the two-photon Rabi 
frequency coupling between atomic states $g$ and $g'$.   
Typically $|\vec k_{1}|\approx |\vec k_{2}|$ and we can use 
$\eta_{0 m_b}=\langle 0|\cos({\vec k_{1}\cdot\vec
r})|m_b\rangle$ for both free-bound transition amplitudes. 
$|\eta_{0 m_b}|^{2}$ is essentially the Franck-Condon factor, or 
a measure of the strength of the 
free-bound transition and will vary considerably with
$\vec r$ for small values of $r$, due to the relatively short
range of the molecular interactions. It is therefore
important to pick the intermediate state $|m_{b} \rangle$ which 
results in a large $|\eta_{0 m_b}|$.
These values can be determined for selected molecular
states from the results of photo-association experiments. They can
also be computed directly if accurate potentials are available.
Detailed discussions are given in~\cite{verhaar,julienne,matt}.

For two counter-propagating waves 
($\vec K/2\approx \vec k_{1}\approx -\vec k_{2}$) 
and $\mu =g$, the above result is
similar to the single atom case, which is often referred to as 
Bragg diffraction~\cite{nist}. The coupling involves the
simultaneous absorption and stimulated emission of photons, i.e.
the elementary process involves a total of two photons. In
contrast to the single atom case, a pair of atoms are
now involved. Hence for Bose condensed atoms with $p_i\approx 0$,
Bragg diffraction produces pairs of atoms in the state
$\alpha|p_1\approx 0,p_2\approx
0\rangle +\beta|\vec p_1=\hbar\vec K/2,\vec
p_2=\hbar\vec K/2\rangle$. 
The momentum shift per atom is only half the
value of atomic Bragg diffraction, due to the fact that only
two photons are involved for the pair of atoms.
Hence the resonance (energy conservation) condition $\Delta\omega$ 
occurs at half the atomic Bragg resonance.

For a Raman process with two co-propagating waves 
($\vec k_{1}\approx \vec k_{2}, \vec K/2\approx 0$)
between two nearly degenerate states $g$ and $g'$
(as depicted in Fig.\ref{fig1}(b)), 
the resonance condition is
$\Delta\omega=2\omega_{g'g}$, twice the atomic resonance; 
i.e. the Raman process changes
each atom's state ($g\rightarrow g'$) and so each atom acquires
the energy deficit $\hbar\omega_{g'g}$. Note that, in this case, there is no
$\vec R$ dependence in $V_{R}$.

We now consider many atoms in a spin 1/2 system to investigate 
the extent to which Raman coupling via an intermediate molecular state 
can produce spin squeezing~\cite{ueda} and correspondingly massive 
entanglement~\cite{zoller}. If we designate $|g\rangle$ and $|g'\rangle$ as
$|\!\uparrow\rangle$ and $|\!\downarrow\rangle$ 
for spin up and down, respectively, our Raman coupling 
for the two particle case is of the form
\begin{eqnarray}
&&{\hbar\Omega_R\over 2}
\left[(|\!\uparrow\rangle\!\langle\downarrow\!|)_1\otimes
(|\!\uparrow\rangle\!\langle\downarrow\!|)_2
+(|\!\downarrow\rangle\!\langle\uparrow\!|)_1\!\otimes\!
(|\!\downarrow\rangle\!\langle\uparrow\!|)_2\right]\nonumber\\
=&&{\hbar\Omega_R\over 2}(\sigma_x^{(1)}\!\otimes\!\sigma_x^{(2)}
-\sigma_y^{(1)}\!\otimes\!\sigma_y^{(2)}),
\end{eqnarray}
where $\sigma_{\mu=x,y,z}$ are the Pauli matrices.  
This is a new type of coupling not widely
discussed before. Its ability, however, to generate entanglement
should be obvious as two atoms perform conditional evolution at
all times.

We first distinguish our coupling scheme from other relevant models.
In the original scheme of M\o lmer and S\o rensen~\cite{mol},
the two atom coupling takes the form
${1\over 2}{\hbar\Omega_R}\sigma_x^{(1)}\!\otimes\!\sigma_x^{(2)}$,
a form different from ours. If the pair-wise 
interaction acts indiscriminately for all pairs of atoms, it is convenient to
analyze the effect of such couplings for many atoms in terms of a collective
spin operator $J_{\mu}=\sum_{i}\sigma_{\mu}^{(i)}, \mu = x,y,z$, 
where the sum is over the number of atoms $N$~\cite{sipe,milburn,chris,yurke}.
One can then show that the Hamiltonian for the
M\o lmer and S\o rensen scheme becomes
$V_{M}=\sum_{i<j}{1\over 2}{\hbar\Omega_R}\sigma_x^{(i)}\!\otimes\!\sigma_x^{(j)}
={1\over 4}{\hbar\Omega_R}(J_x^2-N)$.
In contrast, our scheme gives
\begin{eqnarray}
\sum_{i<j}{\hbar\Omega_R\over 2}(\sigma_+^{(i)}\!\otimes\!\sigma_+^{(j)}
+\sigma_-^{(i)}\!\otimes\!\sigma_-^{(j)})={\hbar\Omega_R\over 2}(J_x^2-J_y^2).
\label{ourscheme}
\end{eqnarray}

Recently, a many body, two mode coupling scheme 
was proposed by S\o rensen {\it et al.}~\cite{zoller}. 
They considered a two
component (i.e. $|\!\downarrow\rangle$ and $|\!\uparrow\rangle$) 
condensate weakly interacting via s-wave collisions described by a 
mean-field. In the approximation where each component has the same spatial
mode, the interaction in terms of the collective spin operators 
$J_{x,y,z}$ is of the form $V_{S}={1\over 2}\hbar\Omega_R J_z^2$.

In order to compare the various coupling schemes, we investigate 
the time evolution assuming a pure initial state with a 
fixed, total number of atoms $N$.  Using the notation of second 
quantization, where $a_j^\dag$ and $a_j$ are the operators for creating and 
annihilating a particle in state $j$ ($j\!=\uparrow,\downarrow$), the 
collective spin operators can be written as 
$J_x=(a_{\downarrow}^\dag a_{\uparrow}+a_{\uparrow}^\dag a_{\downarrow})/2$,
$J_y=i(a_{\downarrow}^\dag a_{\uparrow}-a_{\uparrow}^\dag a_{\downarrow})/2$ and
$J_z=(a_{\uparrow}^\dag a_{\uparrow}-a_{\downarrow}^\dag a_{\downarrow})/2$. 
To numerically calculate the time evolution, 
we expand the wave function as
$|\psi\rangle=\sum_{m=0}^{N} c_m(t) |m\rangle_{\downarrow}|N-m\rangle_{\uparrow}$,
where $|m\rangle_j={(a_j^{\dag})^m}|{\rm vac}\rangle/\sqrt{m!}$,
with the initial conditions given by the $c_m(0)$s.

\begin{figure}
\vskip -0.05in
\centerline{\psfig{figure=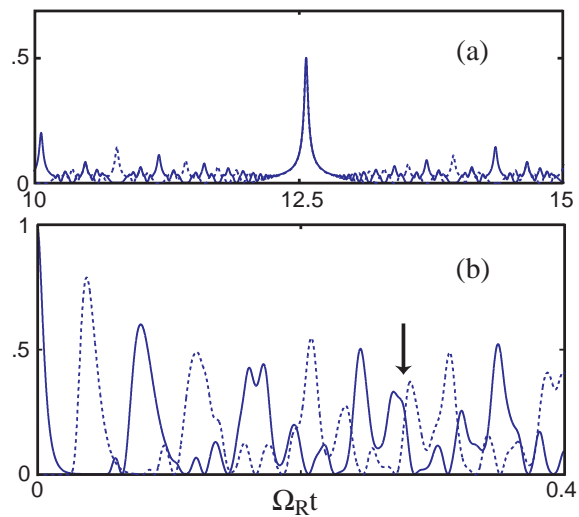,width=3.0in}\\[12pt]}
\vskip -0.125in
\caption{The time dependent probabilities $|c_0|^2$ (solid line) 
and $|c_N|^2$ (dashed line)
of being in states $|N\rangle_{\downarrow}|0\rangle_{\uparrow}$
and $|0\rangle_{\downarrow}|N\rangle_{\uparrow}$, respectively,
for (a) the M\o lmer and S\o renson coupling ($V_{M}$) and (b) our
``molecular Raman" coupling (Eq.\ref{ourscheme}).  The initial
conditions are $c_m(0)=\delta_{0m}$. The number of atoms in the
simulation is $N=10^3$. The projection onto the
$|{\rm GHZ}\rangle_N$ state corresponds to
$|c_0|^2=|c_N|^2$. In (a) it is
apparent that the coupling $V_{M}$ produces a perfect $|{\rm
GHZ}\rangle_N$ state at $\Omega_R t=4\pi$. While in (b) it
appears that at $\Omega_R t\approx 0.28$ (indicted by the arrow) 
our coupling scheme results in about 50\% of the atoms 
in the $|{\rm GHZ}\rangle_N$ state.} 
\label{fig2}
\end{figure}

For $N$ an even integer~\cite{even}, the time evolution operator for $V_M$ takes a simple, 
analytical form at $\Omega_Rt={\pi/2}$,
producing a massive GHZ-type wave function
$|{\rm GHZ}\rangle_N=[|N\rangle_{\uparrow}|0\rangle_{\downarrow}+
\eta |0\rangle_{\uparrow}|N\rangle_{\downarrow}]/{\sqrt 2}$,
where $\eta$ is purely a phase factor.
In the context of Bose-Einstein condensation, this is an example
of an interesting fragmented condensate~\cite{james}.
While the M\o lmer and S\o rensen coupling ($V_{M}$) produces perfect
GHZ-type states at selected times, the S\o rensen {\it et al.} spin squeezing 
scheme ($V_{S}$) and our scheme (Eq.(\ref{ourscheme})), in
general, do not produce exact GHZ-type states. From numerical
simulations, however, we find that our model can produce
more than 50\% overlap with GHZ-type state $|{\rm GHZ}\rangle_{N}$ at
selected times (see Fig. \ref{fig2}).

We can also compare the achievable spin squeezing
between our scheme (Eq.(\ref{ourscheme})) and that of 
S\o rensen {\it et al.} ($V_{S}$),
using the squeezing parameter
$\xi^2={N(\Delta J_{\vec n_1})^2/
(\langle J_{\vec n_2}\rangle^2+\langle J_{\vec n_3}\rangle^2)}$,
where ${\vec n_i}, i=1,2,3$ are mutually orthogonal 
unit vectors~\cite{zoller,ueda}.

S\o rensen {\it et al.}~\cite{zoller}
have shown that $\xi^2(t>0)<1$ for some
set of $\vec n_i$'s.
Their scheme ($V_{S}$) is in fact the one-axis twisting
model considered by Kitagawa and Ueda earlier~\cite{ueda}.
In this case the problem can be solved analytically, with the
result $(\Delta J_{\vec n_1}^{2})_{\rm min}\sim N^{1/3}$.

On the other hand, our coupling resembles the
the two-axis countertwisting model of Kitagawa and Ueda~\cite{ueda},
and has to be solved numerically. In the limit of
large $N$ and with the condensate initially in one spin state, $J_z=-N/2$,
one can show that
$(\Delta J_{\vec n_1}^{2})_{\rm min}\approx {1/2}$.
The optimal squeezing in this case occurs along $\hat{x}+\hat{y}$. This
result can be easily verified by making a semi-classical
approximation in the dynamical equations for $J_x$ and $J_y$. We
find that the time scale of reaching maximum squeezing is $\sim
1/(N\Omega_R)$ (see also~\cite{wineland}). For condensates
containing $10^6$ atoms, even with a very weak coupling
$\Omega_R\approx 1$ (Hz), the maximum squeezing is reached within a
microsecond.

\begin{figure}
\vskip -0.05in
\centerline{\psfig{figure=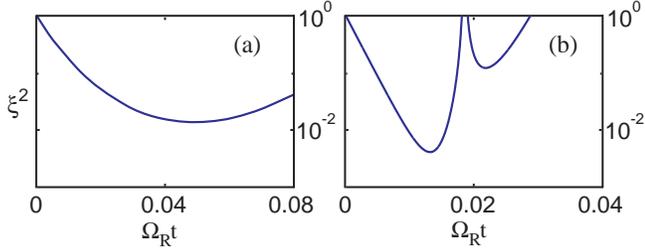,width=3.35in}\\[12pt]}
\vskip -0.125in
\caption{The time evolution of the minimum spin squeezing
parameter $\xi^2$ for (a) the coupling $V_{S}$ 
and (b) the coupling given by Eq.(\ref{ourscheme}).}
\label{fig3}
\end{figure}

In Figure \ref{fig3}, we show the numerically computed, 
minimum squeezing parameters, $\xi^2$, as a function of time
for the coupling from our scheme (Eq.(\ref{ourscheme})) and from the 
S\o rensen {\it et al.} scheme ($V_{S}$). 
In contrast to the S\o rensen {\it et al.} 
scheme, our scheme achieves better squeezing
at an earlier time. In addition for $V_{S}$, the direction $\vec 
n_{1}$ along which minimum squeezing is observed 
varies with time~\cite{zoller,ueda}. 
While for the coupling of Eq.(\ref{ourscheme}), 
it is fixed along $\hat{x}+\hat{y}$~\cite{ueda}.

In order to realize our coupling scheme 
experimentally, the molecular coupling (Fig.1(b), right-side) needs to dominate 
over the atomic coupling (Fig.1(b), left-side). 
That is, we need to achieve a molecular coupling
$\Omega_R^{M}={\Omega_1\Omega_2^*}|\eta_{0m_b}|^{2}/\Delta_M
\gg\Omega_R^{A}={\Omega_1\Omega_2^*}/\Delta_A$,
where $\Delta_{M/A}$ are detunings from the molecular
and atomic intermediate states, respectively.
We should also have $\Delta_M\!\gg\!\gamma_M$, the excited 
molecular state linewidth, to minimize
spontaneous emission which would lead to decoherence and loss of atoms.
These constraints suggest that a deep
molecular bound state with significant transition strength should be chosen in
order to maximize the detuning from the atomic transition and achieve 
a sufficient detuning from excited molecular states. In addition, 
if $\Delta_A\!\gg\!\Delta_M$ further suppression of the atomic transition 
might be possible by a suitable choice of laser polarizations.

Another mechanism for suppressing the atomic coupling with respect to the 
molecular coupling is the two-photon, Raman resonance detuning. 
For Bragg diffraction, the transition via a molecular 
coupling occurs at half the detuning of the transition via an atomic 
coupling (50 kHz vs. 100 kHz, respectively for sodium).  
If the Raman transition involves changes in the internal state of the 
atom, then the frequency for the Raman transition via a molecular 
coupling will be at twice the frequency of the Raman transition via 
an atomic coupling (see Fig. 1(b)).  For Raman transitions between 
Zeeman sublevels in modest magnetic fields this frequency difference 
can be several MHz, which would greatly suppress the single atom 
transition.  

There are several advantages of our coupling scheme. 
The different Raman resonance frequency is a clear signature 
for the Raman transition via a molecular intermediate state.  
In addition, for Bragg diffraction via the molecular coupling, the atoms would
move at half the speed of atoms that have undergone atomic Bragg
diffraction. 

Our scheme is based on an engineered interaction that can be
turned on and off like the scheme of M\o lmer and S\o rensen 
($V_{M}$). In contrast to the S\o rensen {\it et al.} coupling scheme ($V_{S}$), 
our scheme works for non-interacting ground state atoms, 
which can decrease the noise due to
atom-atom interactions in a $U(1)$ symmetry breaking condensate
state. Our scheme also achieves the same level of squeezing 
and the same high value of overlap with the massive GHZ state, 
even as the number of atoms is increased.

Finally, we note that our discussions above can also be applied to
a pair of different species of atoms. For example, Raman coupling using a molecular
intermediate state of the Li-Cs dimer could create entangled pairs
of Li and Cs. 

We thank M. S. Chapman, S. L. Rolston and A. S\o rensen for helpful
discussion. K.H. acknowledges support in part from the Office of 
Naval Research, ARDA and NASA. L.Y. acknowledges support through ONR grant No.
14-97-1-0633 and ARO/NSA grant G-41-Z05.


\begin{references}
\bibitem{lowenergy}
We note that
if the range of the inter-atomic interaction
$\ll r\ll $ size of the ground state
of the harmonic trap, then these bound
states should resemble the standard low energy
scattering solutions.
\bibitem{secondorder}
To be formally correct, the second
order perturbation process also needs to be considered for the
state $|g',g\rangle$, which should have some type of effective
couplings with itself and as well as with the states $|g,g\rangle$
and $|g',g'\rangle$. We have neglected those couplings, assuming
that resonance conditions for their survival in the interaction
picture are not satisfied.
\bibitem{even}
We see that for all these models, there exists an even/odd
number problem: The final state depends
on whether the total number of condensed
atoms is even or odd number. If, however, the condensate
resembles a coherent superposition of number states (grand canonical ensembles),
we should average the results over the number distribution
$p(N)$ as well.
\end{references}
\end{document}